# PARALLEL ROUTING IN MOBILE AD-HOC NETWORKS

Khaled Day, Abderezak Touzene, Bassel Arafeh, Nasser Alzeidi
Department of Computer Science, Sultan Qaboos University, Muscat, Oman
kday@squ.edu.om, arafeh@squ.edu.om, touzene@squ.edu.om,
alzidi@squ.edu.om

*ABSTRACT*

*This paper proposes and evaluates a new position-based Parallel Routing Protocol (PRP) for simultaneously routing multiple data packets over disjoint paths in a mobile ad-hoc network (MANET) for higher reliability and reduced communication delays. PRP views the geographical region where the MANET is located as a virtual 2-dimensional grid of cells. Cell-disjoint (parallel) paths between grid cells are constructed and used for building pre-computed routing tables. A single gateway node in each grid cell handles routing through that grid cell reducing routing overheads. Each node maintains updated information about its own location in the virtual grid using GPS. Nodes also keep track of the location of other nodes using a new proposed cell-based broadcasting algorithm. Nodes exchange energy level information with neighbors allowing energy-aware selection of the gateway nodes. Performance evaluation results have been derived showing the attractiveness of the proposed parallel routing protocol from different respects including low communication delays, high packet delivery ratios, high routing path stability, and low routing overheads.*

## 1. INTRODUCTION

Communication in a Mobile Ad-hoc Network (MANET) is a challenging problem due to node mobility and energy constraints. Many routing protocols for MANETs have been proposed which can be broadly classified in two categories: *topology-based routing* and *position-based routing*. In topology-based protocols [1], link information is used to make routing decisions. They are further divided in: proactive protocols, reactive protocols and hybrid protocols, based on when and how the routes are discovered. In proactive topology-based protocols, such as DSDV [2], each node maintains tables containing routing information to other nodes in the network. When the network topology changes the nodes propagate update messages throughout the network to maintain a consistent and up-to-date view of the network [3]. In reactive topology-based protocols, such as AODV [4], the routes are created only when needed. Hybrid protocols, such as: ZRP [5], combine both proactive and reactive approaches where the nodes proactively maintain routes to nearby nodes and establish routes to far away nodes only when needed.

The second broad category of routing protocols is the class of position-based protocols [6-9]. These protocols make use of the nodes' geographical positions to make routing decisions.

Nodes are able to obtain their own and destination's geographical position via Global Positioning System (GPS) and location services. This approach has become practical by the rapid development of hardware and software solutions for determining absolute or relative node positions in MANETs [10]. One advantage of this approach is that it requires limited or no routing path establishment/maintenance which constitutes a major overhead in topology based routing methods. Another advantage is scalability. It has been shown that topology based protocols are less scalable than position-based protocols [6]. Examples of position-based routing algorithms include: POSANT (Position Based Ant Colony Routing Algorithm) [9], BLR (Beaconless Routing Algorithm) [11], and PAGR (Power Adjusted Greedy Routing) [12].

In [13], a location aware routing protocol (called GRID) for mobile ad-hoc networks was proposed. GRID views the geographic area of the MANET as a virtual 2D grid with an elected leader node in each grid square. Routing is performed in a square-by-square manner through the leader nodes. The GRID protocol involves a number of costly overhead procedures including: leader election, route search and route reply procedures, route maintenance, and a handoff procedure involving broadcast of the routing table when a leader moves away from its grid cell. Variants of GRID have been proposed in [14] and [15] introducing some improvements. In [14] nodes can enter a sleep mode to conserve energy and in [15] stable gateway nodes that stay as long as possible in the same cell are selected. Other protocols based on a virtual grid view have appeared such as those reported in [16] and [17].

In this paper we propose a parallel routing protocol (called PRP) that uses a similar two-dimensional grid view of the MANET geographic area. It allows a source node $S$ to send to a destination node $D$ a set $P = \{p_1, p_2, …, p_m\}$ of $m$ packets in parallel over $n$ disjoint paths, $1 \leq n \leq 8$. The $m$ packets could correspond to the same packet sent in multiple copies for reliability or could be parts of divided up large data sent in parallel for faster delivery. A number of challenges need to be addressed in designing such a protocol including: dynamic topology, node mobility, energy-constrained operation, limited bandwidth, and limited radio transmission ranges.

The proposed PRP uses a virtual 2-dimensional grid view of the geographical region. It is built around three components: parallel path construction, node positioning, and gateway selection. Only one selected gateway node in each grid cell contributes to the forwarding of packets through that cell reducing the communication overhead. Nodes exchange with their neighboring nodes periodic beacon packets carrying residual energy information allowing energy-aware selection of gateway nodes. Each node keeps track of its own location in the virtual grid using GPS positioning. Nodes also keep track of each other's locations in the virtual grid by broadcasting special cell-exit control packets when a node moves out of its current cell to a neighboring cell. For this purpose we propose an efficient new cell-based broadcasting algorithm that propagates the cell-exit packets only via a subset of gateway nodes.

The remainder of the paper is organized as follows: section 2 introduces notations and assumptions; section 3 describes routing path construction; section 4 describes node positioning mechanisms; section 5 explains the gateway selection method; section 6 outlines the PRP routing algorithm; section 7 presents performance evaluation results; and section 8 concludes the paper.

## 2. NOTATIONS AND ASSUMPTIONS

Consider a mobile ad hoc network (MANET) composed of *N* mobile wireless devices (nodes) distributed in a given geographical region. We want to design a parallel routing protocol PRP that allows any MANET source node *S* to send to any MANET destination node *D* a set $P = \{p_1, p_2, …, p_m\}$ of *m* packets in parallel over *n* cell-disjoint paths with the largest possible *n*. The *m* packets could correspond to the same packet sent in multiple copies in which case the objective would be to increase reliability. The *m* packets could alternatively correspond to pieces of a divided up large data message sent in parallel in order to reduce the message transmission delay.

The proposed PRP solution to this problem views the geographical region where the MANET nodes are located as a virtual $k \times k$ two-dimensional (2D) grid of cells as shown in figure 1. The length of a side of a grid cell is denoted *d*. Two grid cells are called neighbor cells if they have a common side or a common corner. Therefore each grid cell has eight neighbor cells. A path in the 2D-grid is a sequence of neighboring grid cells.

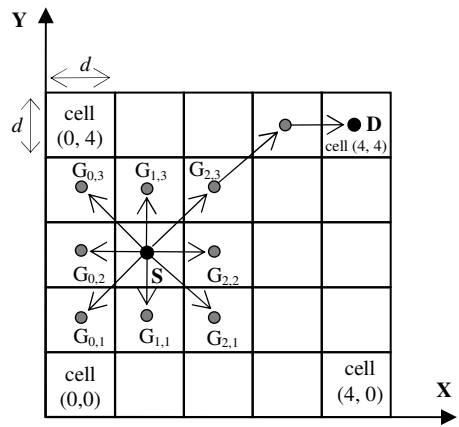

S: source node, D: destination node, $G_{x,y}$: gateway node in cell (*x*,*y*)

Figure 1: Virtual 2D Grid View of the Geographical Area of a MANET

Two nodes are called neighbor nodes if they are located in neighbor cells. The value of *d* is selected depending on the transmission range *r* (all nodes are assumed to have the same transmission range *r*) such that each node can communicate directly with all nodes located in neighboring grid cells. This requirement is met if *d* satisfies the condition $r \geq 2d\sqrt{2}$. This can be seen by noticing that the farthest apart points in two neighboring grid cells are two diametrically opposite corners at distance 2*d* in each of the two dimensions. These two farthest apart points are at distance: $2\sqrt{2}d$. Each grid cell is identified by a pair of integer grid coordinates (*x*, *y*) as illustrated in figure 1. Each node has a distinctive node id (IP or MAC address). Letters such as *A*, *B*, *S*, *D* and *G* are used to represent node ids. A packet sent by a node is of one of three types: a unicast packet to be received by a single node within the sender's transmission range (namely the node whose node id is indicated in the receiver address field of the packet); a local broadcast packet to be received by all nodes in neighboring grid cells or a global broadcast packet to be forwarded to all nodes based on the broadcasting algorithm rules described later in the paper.

The proposed PRP protocol routes packets in parallel over multiple cell-disjoint paths. Along each path, a data packet is routed from a source node *S* to a destination node *D* over the 2D-grid structure with each routing step moving the packet from a node at a grid cell to a selected gateway node in a neighboring grid cell until the destination node *D* is reached. Each node *S* maintains a list of gateway nodes located in the neighboring grid cells. Only gateway nodes participate in the forwarding of packets through the sequence of cells that forms a routing path. A gateway node in cell ($x$, $y$) is denoted $G_{x,y}$. Each node can have up to 8 neighboring gateway nodes (one in each of the 8 neighboring grid cells) as illustrated in figure 1. We assume each node is able to obtain its own geographical position (through a GPS receiver) and map it to a pair of integer cell coordinates. We also assume each node is able to read its energy level at any time.

## 3. CONSTRUCTION OF CELL-DISJOINT (PARALLEL) PATHS IN A 2D-GRID

Let *S* be a source node located in a source cell ($x_S$, $y_S$) of a virtual 2D-grid and let *D* be a destination node located in a destination cell ($x_D$, $y_D$). We show how to construct a maximum number of cell-disjoint paths from *S* to *D*. A path from *S* to *D* is a sequence of cells starting with the source cell ($x_S$, $y_S$) and ending with the destination cell ($x_D$, $y_D$) such that any two consecutive cells in the sequence are neighbor grid cells. Two paths from *S* to *D* are called cell-disjoint if they do not have any common cells other than the source cell ($x_S$, $y_S$) and the destination cell ($x_D$, $y_D$). A path from *S* to *D* can be defined by the sequence of cell-to-cell moves that lead from *S* to *D*.

There are eight possible moves from any cell to a neighbor cell. These eight moves are denoted: <+x>, <-x>, <+y>, <-y>, <+x, +y>, <+x, -y>, <-x, +y> and <-x, -y>. The moves <+x> and <-x> correspond to the right and left horizontal moves in the grid, the moves <+y> and <-y> correspond to the up and down vertical moves and the moves <+x, +y>, <+x, -y>, <-x, +y> and <-x, -y> correspond to the four diagonal moves right-up, right-down, left-up, and left-down. In a path description we use a superscript value *i* after a move to represent *i* consecutive repetitions of that move. For example <+x, +y>$^3$ represents 3 successive right-up diagonal moves.

Table 1: Cell-Disjoint Paths for the case $\delta_x > \delta_y \geq 1$

| Path | Source Exit Moves | Diagonal Moves | Horizontal Moves | Destination Entry Moves |
|---|---|---|---|---|
| $\pi_{11}$ | <+x, +y> | <+x, +y>$^{\delta_y-1}$ | <+x>$^{\delta_x-\delta_y-1}$ | <+x> |
| $\pi_{12}$ | <+x> | <+x, +y>$^{\delta_y-1}$ | <+x>$^{\delta_x-\delta_y-1}$ | <+x, +y> |
| $\pi_{13}$ | <+y> <+x, +y> | <+x, +y>$^{\delta_y-1}$ | <+x>$^{\delta_x-\delta_y-1}$ | <+x, -y> |
| $\pi_{14}$ | <+x, -y> | <+x, +y>$^{\delta_y-1}$ | <+x>$^{\delta_x-\delta_y-1}$ | <+x, +y> <+y> |
| $\pi_{15}$ | <-x, +y> <+x, +y>$^2$ | <+x, +y>$^{\delta_y-1}$ | <+x>$^{\delta_x-\delta_y-1}$ | <+x, -y><-y> |
| $\pi_{16}$ | <-y> <+x, -y> | <+x, +y>$^{\delta_y-1}$ | <+x>$^{\delta_x-\delta_y-1}$ | <+x, +y>$^2$ <-x, +y> |
| $\pi_{17}$ | <-x> <-x, +y> <+x, +y>$^3$ | <+x, +y>$^{\delta_y-1}$ | <+x>$^{\delta_x-\delta_y-1}$ | <+x, -y>$^2$ <-x, -y> |
| $\pi_{18}$ | <-x, -y> <+x, -y>$^2$ | <+x, +y>$^{\delta_y-1}$ | <+x>$^{\delta_x-\delta_y-1}$ | <+x, +y>$^3$ <-x, +y> <-x> |

There are at most eight cell-disjoint paths from *S* to *D* corresponding to the eight possible starting moves: <+x>, <-x>, <+y>, <-y>, <+x, +y>, <+x, -y>, <-x, +y>, and <-x, -y>. We assume without loss of generality that $x_D \geq x_S$ and $y_D \geq y_S$ (i.e. *D* right-up from *S*). The paths for the other conditions can be derived from the paths of the condition $x_D \geq x_S$ and $y_D \geq y_S$ as follows: (a) if $x_D \geq x_S$ and $y_D < y_S$ then replace $+y$ by $-y$ and vice versa in all paths, (b) if $x_D < x_S$ and $y_D \geq y_S$ then replace $+x$ by $-x$ and vice versa in all paths, and (c) if $x_D < x_S$ and $y_D < y_S$ then do both replacements in all paths. We distinguish four cases in the construction of cell-disjoint paths from the source cell $(x_S, y_S)$ to the destination cell $(x_D, y_D)$ depending on the relationship between the distances $\delta_x = x_D - x_S$ and $\delta_y = y_D - y_S$ along the $x$ and $y$ dimensions.

*Case 1: If $\delta_x > \delta_y \geq 1$* (the case $\delta_y > \delta_x \geq 1$ is symmetric, it can be obtained by swapping $x$ and $y$):

Table1 lists 8 cell-disjoint paths from source cell $(x_S, y_S)$ to destination cell $(x_D, y_D)$ for this case.

Each path starts with a sequence of *source exit moves* which include a move to one of the 8 neighbor cells of the source cell followed by up to four moves to reach the common exit column (the grid column after the source column in the direction of the destination column, see figure 2).

Notice that in the symmetric case $\delta_y > \delta_x \geq 1$, *column* should be replaced by *row*. Once the paths reach the exit column they all follow the same two sequences of moves which are a sequence of $\delta_y-1$ diagonal moves of the type $<+x, +y>$ followed by a sequence of $\delta_x-\delta_y-1$ horizontal moves of the type $<+x>$. In the symmetric case $\delta_y > \delta_x \geq 1$, *horizontal* should be replaced by *vertical*.

These two sequences make the eight paths reach the destination entry column which is the column immediately preceding the destination column. Once the entry column is reached the paths follow a sequence of up to five *destination entry moves* to maintain the cell-disjoint property. Figure 2 illustrates the construction of this case for $\delta_x = 5$ and $\delta_y = 2$.

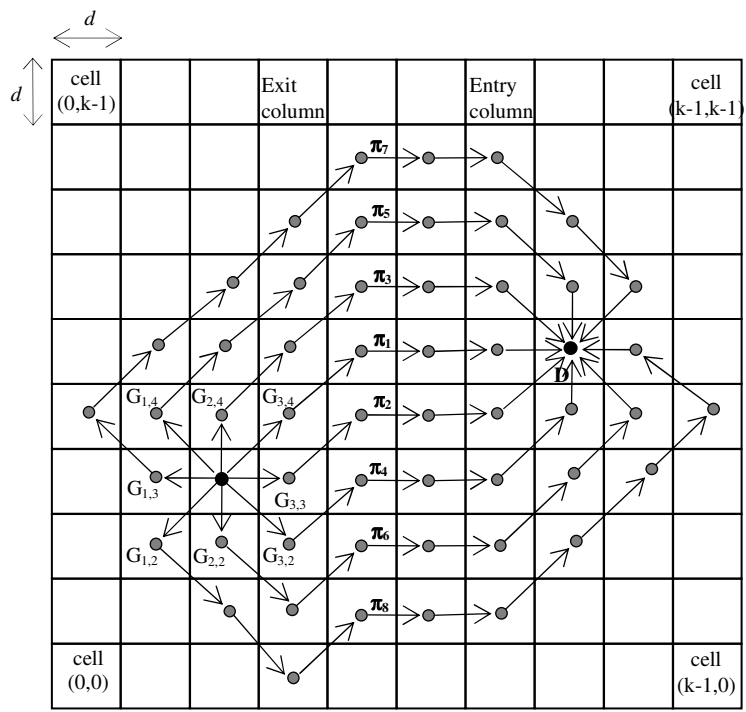

Figure 2: Cell-Disjoint Paths in a 2D Grid

*Case 2:* $\delta_x \geq 2$ & $\delta_y = 0$ (the case $\delta_y \geq 2$ & $\delta_x = 0$ can be obtained by swapping *x* and *y*):

Table 2 lists 8 cell-disjoint paths from source cell $(x_S, y_S)$ to destination cell $(x_D, y_D)$ for this case.

Table 2: Cell-Disjoint Paths for the case $\delta_x \geq 2$ and $\delta_y = 0$

| Path | Source Exit Moves | Horizontal Moves | Destination Entry Moves |
|---|---|---|---|
| $\pi_{21}$ | <+x> | <+x>$^{\delta_x - 2}$ | <+x> |
| $\pi_{22}$ | <+x, +y> | <+x>$^{\delta_x - 2}$ | <+x, -y> |
| $\pi_{23}$ | <+x, -y> | <+x>$^{\delta_x - 2}$ | <+x, +y> |
| $\pi_{24}$ | <+y> <+x, +y> | <+x>$^{\delta_x - 2}$ | <+x, -y> <-y> |
| $\pi_{25}$ | <-y> <+x, -y> | <+x>$^{\delta_x - 2}$ | <+x, +y> <+y> |
| $\pi_{26}$ | <-x, +y> <+x, +y>$^2$ | <+x>$^{\delta_x - 2}$ | <+x, -y>$^2$ <-x, -y> |
| $\pi_{27}$ | <-x, -y> <+x, -y>$^2$ | <+x>$^{\delta_x - 2}$ | <+x, +y>$^2$ <-x, +y> |
| $\pi_{28}$ | <-x> <-x, +y> <+x, +y>$^3$ | <+x>$^{\delta_x - 2}$ | <+x, -y>$^3$ <-x, -y> <-x> |

*Case 3:* $\delta_x = 1$, $\delta_y = 0$ (the case $\delta_y = 1$, $\delta_x = 0$ is symmetric)

Table 3 lists eight parallel paths from cell $(x_S, y_S)$ to cell $(x_D, y_D)$ for this case.

Table 3: Cell-Disjoint Paths for the case $\delta_x = 1$ and $\delta_y = 0$

| Path | Source Exit Moves | Destination Entry Moves |
|---|---|---|
| $\pi_{31}$ | <+x> | |
| $\pi_{32}$ | <+x, +y> | <-y> |
| $\pi_{33}$ | <+x, -y> | <+y> |
| $\pi_{34}$ | <+y> | <+x, -y> |
| $\pi_{35}$ | <-y> | <+x, +y> |
| $\pi_{36}$ | <-x, +y> <+x, +y> | <+x> <+x, -y> <-x, -y> |
| $\pi_{37}$ | <-x, -y> <+x, -y> | <+x> <+x, +y> <-x, +y> |
| $\pi_{38}$ | <-x> <-x, +y> <+x, +y>$^2$ | <+x> <+x, -y>$^2$ <-x, -y> <-x> |

*Case 4:* $\delta_x = \delta_y$ (notice that we must have here $\delta_x = \delta_y \geq 1$)

Table 4 lists eight parallel paths from cell $(x_S, y_S)$ to cell $(x_D, y_D)$ for this case.

Table 4: Cell-Disjoint Paths for the case $\delta_x = \delta_y$

| Path | Source Exit Moves | Common Diagonal Moves | Destination Entry Moves |
|---|---|---|---|
| $\pi_{41}$ | <+x, +y> | <+x, +y>$^{\delta y-1}$ | |
| $\pi_{42}$ | <+x> | <+x, +y>$^{\delta y-1}$ | <+y> |
| $\pi_{43}$ | <+y> | <+x, +y>$^{\delta y-1}$ | <+x> |
| $\pi_{44}$ | <+x, -y> | <+x, +y>$^{\delta y-1}$ | <+x, +y> <-x, +y> |
| $\pi_{45}$ | <-x, +y> <+x, +y> | <+x, +y>$^{\delta y-1}$ | <+x, -y> |
| $\pi_{46}$ | <-y> <+x, -y> | <+x, +y>$^{\delta y-1}$ | <+x, +y>$^2$ <-x, +y> <-x> |
| $\pi_{47}$ | <-x> <-x, +y> <+x, +y>$^2$ | <+x, +y>$^{\delta y-1}$ | <+x, -y> <-y> |
| $\pi_{48}$ | <-x, -y> <+x, -y>$^2$ | <+x, +y>$^{\delta y-1}$ | <+x, +y>$^3$ <-x, +y>$^2$ <-x,-y> |

## 4. NODE POSITIONING

Each node *A* maintains updated information about its location in the virtual grid in the form of a pair of grid coordinates ($x_A$, $y_A$) based on GPS information. Each node *A* also keeps track of the location of other MANET nodes in a table `LOCATION`$_A$`[]` where `LOCATION`$_A$`[`*B*`]` stores at node *A* the pair of grid coordinates ($x_B$, $y_B$) of the cell where node *B* is located. Every time a node *A* detects it has moved to a new grid cell (detection mechanism is assumed in place based on GPS), it initiates the broadcasting of a *cell-exit packet* <A,new_x$_A$,new_y$_A$,seq$_A$> indicating that node *A* has moved to a new grid cell whose coordinates are *new_x$_A$* and *new_y$_A$*. The source relative sequence number `seq`$_A$ helps detecting and discarding duplicate broadcast packets.

```
Cell–Exit (A, new_x_A, new_y_A)
{  // Node A is moving out of its current cell. It updates
   // its cell coordinates record and notifies the other nodes
   o   LOCATION_A[A]= (new_x_A, new_y_A)
   o   seq_A = seq_A + 1
   o   P = <A, new_x_A, new_y_A, seq_A>
   o   send Broadcast(P)
}
```

Figure 3: Cell-Exit Procedure

Figures 3 and 4 outline the main steps of the cell-exit and broadcasting procedures. At most one node in each grid cell (the gateway node of that cell) contributes to the forwarding of broadcast packets. All other nodes receive the broadcast packet, update their records but do not forward it. In fact not all gateway nodes need to forward a broadcast packet. The broadcasting is performed layer-by-layer where layers are co-centric squares centered at the source cell. Only gateways in half (50%) of the cells in each layer forward the broadcast packet. The forwarding and non forwarding gateway nodes alternate in each layer. This is sufficient to ensure all nodes receive the broadcast packet. This broadcasting algorithm is coded in figure 4 and illustrated in figure 5.

```
Broadcast (P = <A, new_x_A, new_y_A, seq_A>)
{  // Node B received a broadcast packet P originated at node A
   o  if ( (A, seq_A) already seen) discard P; exit
   o  LOCATION_B[A]= (new_x_A, new_y_A)
   o  if (B is not a gateway node) exit
   o  if (|x_B-x_A|+|y_B-y_A| is odd) exit //B is a non forwarding gateway
   o  send Broadcast(P) //B is a forwarding gateway
}
```

Figure 4: Cell-Based Broadcasting Algorithm

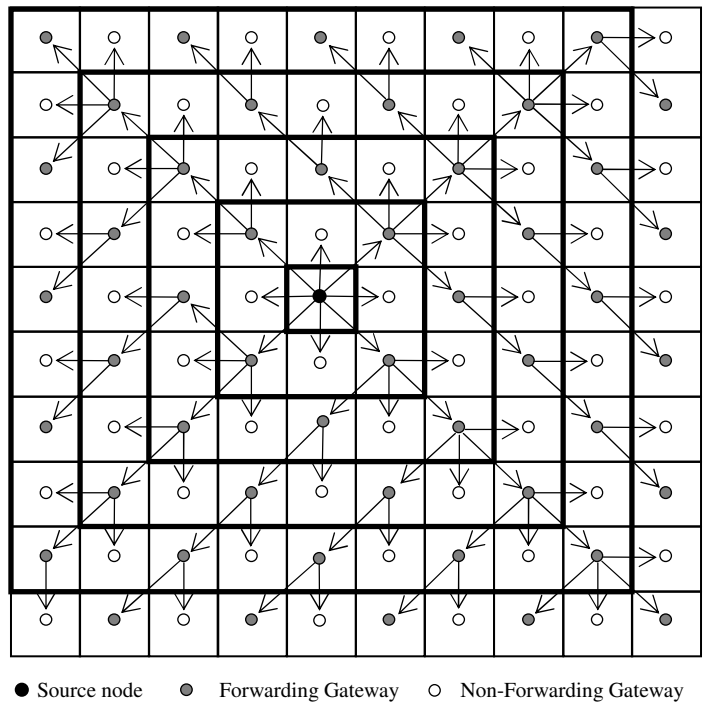

● Source node    ● Forwarding Gateway    ○ Non-Forwarding Gateway

Figure 5: Layer-By-Layer Cell-Based Broadcasting

## 5. GATEWAY SELECTION

Only one selected *gateway* node in each grid cell contributes to the forwarding of unicast and broadcast packets through that grid cell reducing substantially the communication overhead. Nodes exchange with their neighboring nodes periodic *beacon* control packets each containing the sender's cell position and the sender's energy level. When a node *A* receives a *beacon* packet from a node *B* with energy level $E_B$ and located in cell (*x*, *y*), node *A* updates the gateway node $G_{x,y}$ of cell (*x*, *y*) setting it to node *B* if $E_B$ is higher than the previously recorded energy level of $G_{x,y}$. This applies if *A* and *B* are in the same cell or in neighboring cells. Every node will therefore regularly update its record about gateway nodes in all surrounding cells and in the local cell. This also allows each node to know whether it is the current gateway node of the cell where it is located or not and hence whether it is responsible or not for forwarding unicast and broadcast packets. Figure 6 summarizes the gateway update procedure.

```
Gateway Update (A, <B,E_B,x,y>)
{  // Node A invokes this procedure when it receives a beacon message
   // <B,E_B,x,y> from node B with energy E_B located at cell (x,y)
   o  if (x,y)≠(x_A,y_A) and (x,y) is not a neighbor cell of (x_A,y_A)then
         exit; //B is neither in local cell nor in a neighbor cell, ignore
   o  if B ≠ G_x,y and E_B > E_{G_{x,y}} then
         G_x,y = B      //B is selected as the new gateway for cell (x,y)
}
```

Figure 6: Gateway Update Procedure

## 6. THE PRP ROUTING ALGORITHM

The proposed parallel routing algorithm makes use of the three previously described components: cell-disjoint path construction, node positioning, and gateway selection. Each MANET node *A* maintains the following data structures built from these three components:

- Routing Paths: $\pi_i[S, D] = i^{th}$ routing path, $1 \leq i \leq 8$, for source-destination pair (*S*, *D*)
- Node Positions : LOCATION[*B*] = cell at which node *B* is located
- Gateway Nodes: $G_{x,y}$ = gateway node at local or neighboring cell (*x*, *y*)

The routing paths $\pi_i[S, D]$, $1 \leq i \leq 8$, are pre-computed based on the previously described cell-disjoint path construction (see tables 1, 2, 3 and 4) and stored at each node.

```
Source_Route (S, D, p_1, p_2, …, p_m, n)
{  // Source node S wants to send to destination node D the m packets
   // p_1, p_2, …, p_m in parallel over n cell-disjoint paths
   if (n > 8) then n = 8; //maximum number of cell-disjoint paths is 8
   i = 0; j = 0
   while (j < m) //while more packets to send
   {
      o  j = j + 1 // p_j is next packet to send
      o  i = i%n + 1 //  π_i[S, D] is next cell-disjoint path to use
      o  (x, y) = neighbor cell corresponding to first hop of path
            π_i[S,D]
      o  send Gateway_Route(G_{x,y}, p_j, S, D, i) to G_{x,y}
   }
`
```

Figure 7: Routing at a Source Node

Figure 7 outlines the routing function invoked by a source node *S* when it wants to send *m* packets $p_1, p_2, \ldots, p_m$ in parallel to a destination node *D* over *n* cell-disjoint paths, $1 \leq n \leq 8$. The source node scatters the *m* packets over the *n* disjoint paths in a round-robin fashion.

Upon reception of a Gateway_Route packet, a gateway node executes the `Gatway_Route()` function outlined in figure 8.

```
Gateway_Route (G, p, S, D, i)
{   // Gateway node G has received a request to forward packet p initiated
    // at source node S towards destination D over path π_i[S,D]
    o   if (C[D] = C[G])  //destination and G are located in the same cell
        {       deliver p to D
                exit
        }
    o   c = NextCell(C[G], S, D, i) // cell following C[G]on π_i[S,D]
    o   NextG = G_c //gateway node at next cell
    o   send Gateway_Route(NextG, p, S, D, i) to NextG
}
```

Figure 8: Routing at a Gateway Node

The `NextCell()` function uses the pre-computed and stored description of the $i^{th}$ cell-disjoint path $\pi_i[S,D]$ from *S* to *D* and determines the next grid cell the packet has to go to. We can use a simple coding scheme for representing the routing information in a packet. Such a scheme will make it easy to determine the next cell on a routing path. For example we can use a routing descriptor of the form `<case, path, phase, step>` which specifies the case number (1 to 4, corresponding to one of the tables 1 to 4), the path number (1 to 8), the phase number (1 to 4) along the path and the step number in that phase. A `phase` of 1 corresponds to the source exit moves phase (see tables 1-4), a `phase` of 2 corresponds to the diagonal moves phase, a `phase` of 3 corresponds to the horizontal/vertical moves phase and a `phase` of 4 corresponds to the destination entry moves phase. Each phase may involve several moves. The `step` field in the position descriptor indicates the move number in the current phase. Table 5 illustrates how we can use such a coding to build a pre-computed routing table that can be stored at each mobile node. Each packet carries a `<case, path, phase, step>` routing descriptor. A node that needs to route a packet looks up the packet's routing descriptor in the node's routing table to determine the needed move for going to the next grid cell along the path as well as the next values of the phase and step fields of the routing descriptor to be included in the forwarded packet (replacing the previous routing descriptor). The four fields (case, path, phase and step) are small integers and hence consume only few bytes in a packet and in a routing table entry.

Table 5: Routing Table (only entries for path 1 of case 1 are shown for brevity)

| Case | Path | Phase | Step | Next Move | Next Phase | Next Step |
|---|---|---|---|---|---|---|
| 1 | 1 | 1 | 1 | <+x, +y> | 2 | 1 |
| 1 | 1 | 2 | $1 \leq i \leq \delta_y-2$ | <+x, +y> | 2 | $i + 1$ |
| 1 | 1 | 2 | $\delta_y-1$ | <+x, +y> | 3 | 1 |
| 1 | 1 | 3 | $1 \leq i \leq \delta_x-\delta_y-2$ | <+x> | 3 | $i + 1$ |
| 1 | 1 | 3 | $\delta_x-\delta_y-1$ | <+x> | 4 | 1 |
| 1 | 1 | 4 | 1 | <+x> | end of path | - |
| 1 | 1 | end of path | - | - | - | - |

## 7. PERFORMANCE EVALUATION

In this section we derive some performance characteristics of PRP. We first obtain the lengths of the constructed cell-disjoint paths. These lengths are readily obtained from tables 1, 2, 3 and 4.

*Result 1:* The lengths of the constructed 8 paths for each of the 4 cases are listed in Table 6.

Table 6: Lengths of the Parallel Paths

| Path | Case 1 | Case 2 | Case 3 | Case 4 |
|---|---|---|---|---|
| 1 | $\delta_x$ | $\delta_x$ | $\delta_x$ | $\delta_y$ |
| 2 | $\delta_x$ | $\delta_x$ | $\delta_x + 1$ | $\delta_y + 1$ |
| 3 | $\delta_x + 1$ | $\delta_x$ | $\delta_x + 1$ | $\delta_y + 1$ |
| 4 | $\delta_x + 1$ | $\delta_x + 2$ | $\delta_x + 1$ | $\delta_y + 2$ |
| 5 | $\delta_x + 3$ | $\delta_x + 2$ | $\delta_x + 1$ | $\delta_y + 2$ |
| 6 | $\delta_x + 3$ | $\delta_x + 4$ | $\delta_x + 4$ | $\delta_y + 4$ |
| 7 | $\delta_x + 6$ | $\delta_x + 4$ | $\delta_x + 4$ | $\delta_y + 4$ |
| 8 | $\delta_x + 6$ | $\delta_x + 8$ | $\delta_x + 8$ | $\delta_y + 8$ |

We make use of these lengths to derive an upper bound for the average packet delivery probability assuming parallel routing of multiple copies of a packet over the eight disjoint paths.

*Result 2:* In a MANET of $N$ nodes located in a $k \times k$ two-dimensional grid, the average packet delivery probability $P_d$ using PRP satisfies:

$$P_d \geq 1 - [1 - (1 - (1 - 1/k^2)^N)^{k+3}]^8 \qquad (1)$$

*Proof:* A packet will be delivered if at least one of the eight paths is not broken. For a path to be non broken we need to have for each of the grid cells along that path at least one MANET node located in that cell. If in total there are $N$ nodes and if we assume node mobility is such that a node is equally likely to be located in any of the $k^2$ cells at any given time, then the probability that a given node is located in a given grid cell is: $1/k^2$. Hence the probability that a given grid cell does not host any of the $N$ nodes is $P_{empty} = (1 - 1/k^2)^N$. The probability that a given grid cell hosts at least one node is therefore:

$$P_{\text{non empty}} = 1-(1 - 1/k^2)^N \qquad (1.1)$$

The probability that each of the $l$ cells along a path $\pi$ of length $l$ hosts at least one gateway node is:

$$P_{\text{delivery on }\pi} = (1-(1 - 1/k^2)^N)^l \qquad (1.2)$$

This probability decreases as the path length $l$ increases. Let us therefore find an upper bound on the average path length. Based on Table 6, the average increase over the minimum length in the eight routing paths is less than 3 in each of the four cases. It is equal to 2.5 in cases 1, 2 and 3 and it is equal to 2.75 in case 4. The maximum distance between any source cell and any destination cell is $k$ hops ($k$ diagonal moves). Hence the average probability of delivery on a single path satisfies:

$$P_{\text{delivery on a single path}} \geq (1-(1 - 1/k^2)^N)^{k+3} \qquad (1.3)$$

The probability of delivery on at least one of the 8 paths satisfies: $P_d \geq 1-[1-(1-(1 - 1/k^2)^N)^{k+3}]^8$.
QED

The expression of $P_d$ is plotted in figure 9 as a function of the network density $\delta = N/k^2$ which is the average number of nodes per grid cell. The delivery probability approaches 1 when the network density reaches 3 nodes per grid cell corresponding to a density of 3 nodes per $r^2/8$ area.

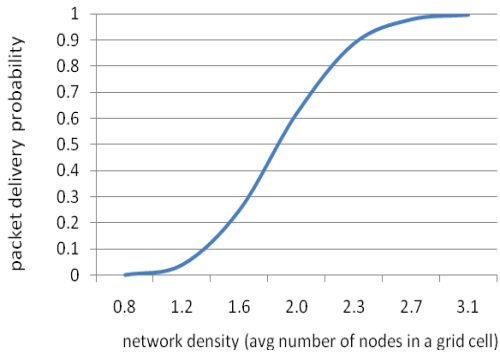

Figure 9: Delivery Probability vs Network Density

Notice that the value of $k$ depends on the size of the physical area and on the transmission range. If we assume a square shaped physical area of size $\Delta$ meters by $\Delta$ meters ($\Delta \times \Delta$ m$^2$), a transmission range of $r$ meters and if we set the grid cell size $d$ to its maximum value $d = r/(2\sqrt{2})$ to reduce the number of hops then the value of $k$ is: $k = \Delta/d = 2\sqrt{2}\Delta/r$. Substituting $k$ by $2\sqrt{2}\Delta/r$ in expression (1) and plotting the packet delivery probability as a function of $r$ for a fixed number of nodes $N = 100$ and a fixed physical area of 500 meters by 500 meters is shown in figure 10. Here we observe the impact of increasing the transmission range which results in reducing *k* and hence the average number of hops in a routing path.

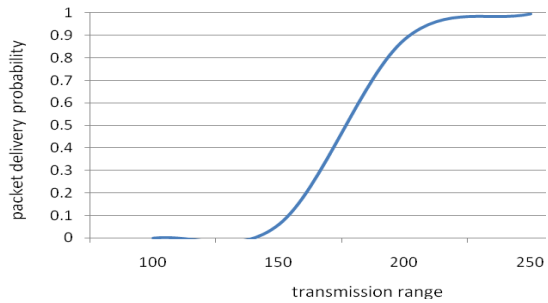

Figure 10: Delivery Probability vs Transmission Range

Now we estimate the total delay to route a large amount of data of size *M* bytes from a source node to a destination node assuming the large amount of data is divided in packets of size *p* bytes each and that these packets are sent in parallel over the cell-disjoint paths. We assume a one-hop packet transmission delay of $\tau$ seconds.

*Result 3:* The delay *T* for sending a large message of size *M* bytes fragmented in packets of size *s* bytes using PRP satisfies: $T \leq M.\tau.(k+8)/8s$, where $\tau$ is the one-hop packet transmission delay.

*Proof:* The total number of packets after fragmenting the message of *B* bytes is: $n = B/s$. Each of the eight cell-disjoint paths between the source and the destination will route $n/8$ of these *n* packets. The total delay on any of the eight parallel paths is at most $(n/8)\tau.l_{max}$, where $l_{max} = k + 8$ is the maximum path length as can be seen in Table 6.  QED

Figure 11 plots the maximum message delay *T* as a function of the transmission range. Higher transmission ranges imply less hops (shorter paths) and hence shorter delays. In this figure we have used the following parameters: $M = 1$ megabytes, $s = 1$ kilobytes and $\tau = 0.008$ seconds. The figure illustrates the reduction in communication delay resulting from sending a large amount of data over the constructed disjoint paths compared to sending it on a single path.

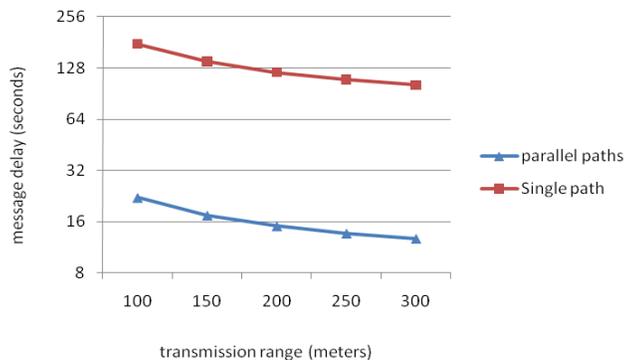

Figure 11: Transmission Delay vs Transmission Range

Now we extend the packet delivery probability expression (1) to take into consideration node mobility.

*Results 4:* The probability of delivery on at least one of the 8 paths between a source cell *SC* and a destination cell *DC* taking into consideration node mobility satisfies:

$$P_d \geq 1-[1-(1-(1-1/k^2)^N)^{k+3}(1-p)^{k+3}]^8. \quad (2)$$

*Proof:* Let *p* be the probability that a mobile node moves out of its current cell after starting transmitting a packet and before completing the transmission of the packet. The distance a mobile node needs to cross to leave its current cell (after starting transmitting a packet) varies between 0 and $d\sqrt{2}$ averaging to $d_{avg} = d\sqrt{2}/2$. If *t* is the packet transmission time and *s* is the node mobility speed, the probability *p* can be estimated as follows (assuming $t \leq \frac{d_{avg}}{s}$):

$$p = \frac{s.t}{d_{avg}} = \frac{2s.t}{d\sqrt{2}} \quad (2.1)$$

For example if *d* = 100m, *s* = 1 m/sec and *t* = 1 sec, then *p* = 0.014.

Let us now revisit the derivation of the packet delivery probability taking into account that a node can move out of its cell while transmitting a packet. The probability that a given grid cell along a routing path of packet hosts at least one (gateway) node and this node does not move out of the cell while forwarding the packet is:

$$P_{\text{non empty cell}} = (1-(1-1/k^2)^N)(1-p) \quad (2.2)$$

The probability that each of the *l* grid cells along a given path $\pi$ of length *l* hosts at least one gateway node and this node does not move out of the cell when forwarding the packet is:

$$P_{\text{delivery on } \pi} = (1-(1-1/k^2)^N)^l (1-p)^l \quad (2.3)$$

As explained in the previous section the average increase over the minimum distance for the eight routing paths does not exceed 3 and the maximum distance (length of a minimum length path) between any source cell and any destination cell is *k* hops yielding an estimated maximum path length equal to *k*+3. Hence the probability of delivery on any routing path $\pi$ satisfies:

$$P_{\text{delivery on } \pi} \geq (1-(1-1/k^2)^N)^{k+3}(1-p)^{k+3} \quad (2.4)$$

The probability of delivery on at least one of the 8 paths between a source cell *SC* and a destination cell *DC* satisfies therefore: $P_d \geq 1-[1-(1-(1-1/k^2)^N)^{k+3}(1-p)^{k+3}]^8$. QED

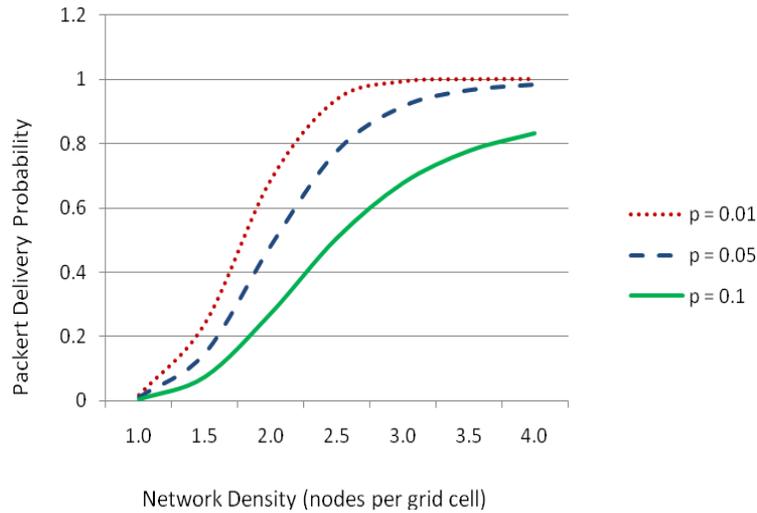

Figure 12: Packet Delivery Probability with Node Mobility

Figure 12 plots the lower bound expression (2) for the packet delivery probability for different values of $p$ and different values of the network density (the network density varies with the number of nodes $N$ and the number of grid cells $k^2$). It can be seen in this figure that even when 10% of the time a gateway node moves out of its cell after starting packet transmission and before completing it (corresponding to $p = 0.1$), packet delivery ratios higher than 0.8 (80%) can still be achieved if the network density is large enough (larger than 3 nodes per cell). Smaller network densities and high mobility (less than 0.05) can result in low packet delivery rates (below 50%). We can conclude that node mobility has limited effect on packet delivery. Finally, we derive a result related to the performance of the cell-based broadcasting algorithm used for node positioning in PRP.

*Result 5:* The number of packets generated by the cell-based broadcasting algorithm used in PRP is $O(\Delta^2/r^2)$ where $\Delta$ is the length (and width) of the MANET physical area and $r$ is the transmission range, independently of the number of nodes in the network. Furthermore the broadcasting delay is $O(\Delta/r)$.

*Proof:* As implemented in figure 4 and illustrated in Figure 5, the cell-based broadcasting algorithm used in PRP requires from only half of the gateway nodes to participate in forwarding a broadcasting packet. In a $k\times k$ grid the number of generated packets is therefore $k^2/2$. If we assume a square shaped physical area of size $\Delta$ meters by $\Delta$ meters ($\Delta\times\Delta$ m$^2$), a transmission range of $r$ meters and if we set the grid cell size $d$ to its maximum value $d = r/(2\sqrt{2})$ to minimize the number of hops then the value of $k$ is: $k = \Delta/d = 2\sqrt{2}\Delta/r$. The number of generated copies of a broadcast packets is therefore $k^2/2 = 4\Delta^2/r^2$ which is $O(\Delta^2/r^2)$ and is independent of the number of nodes in the network. As for the transmission delay, notice that the worst case corresponding to a broadcast packet issues by a source node located in a corner cell in which case the maximum traversed distance by the broadcast packet is $k$ (corner-to-corner diagonal moves). Since $k = 2\sqrt{2}\Delta/r$, the broadcasting delay is $O(\Delta/r)$. QED

Figure 13 shows how the number of packets generated by the cell-based broadcasting algorithm decreases sharply with the increase of transmission range in a 500m×500m network area.

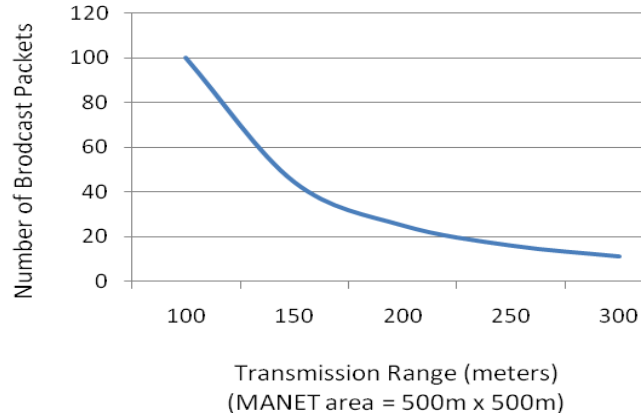

Figure 13: Number of Cell-Based Broadcast Packets vs Transmission Range

## 8. CONCLUSION

This paper has proposed and evaluated a parallel routing protocol (PRP) for mobile ad-hoc networks. PRP allows routing multiple packets in parallel from a source node to a destination node over disjoint paths. Cell-disjoint paths between any two grid cells have been constructed and used for constructing pre-computed routing tables. A single selected (gateway) node in each grid cell contributes to the forwarding of packets through that grid cell reducing substantially the routing overheads. Packets are routed in parallel by the selected gateway nodes along the constructed paths. Each node maintains updated information about its location in the virtual grid using GPS and keeps track of the location of other nodes using a proposed cell-based broadcasting algorithm. Nodes exchange with neighbor nodes energy level information allowing energy-aware selection of the gateway nodes. Performance characteristics of the proposed PRP protocol have been studied. The following can be concluded from the obtained results:

- The lengths (number of hops) of the routing paths used by PRP are within a small additive constant from the optimal length independently of the network size.

- Packet delivery in PRP approaches 100% for non sparse networks.

- Node mobility has limited effect on the performance of PRP due to its cell-based nature. As long as there is at least one node remaining in a cell, the move of other nodes out of the cell does not break the PRP routing paths through that cell.

- The new proposed cell-based broadcasting algorithm used by PRP for propagating node positioning information has a good performance. It generates $O(\Delta^2/r^2)$ messages and incurs an $O(\Delta/r)$ communication delay, where $\Delta \times \Delta$ is the MANET physical area and $r$ is the transmission range.

These results show the attractiveness of PRP as a position-based parallel routing protocol for mobile ad-hoc networks from different respects including low communication delays, high packet delivery ratios, high routing path stability, and low routing overheads.